\documentclass[prl,nofootinbib,twocolumn,superscriptaddress]{revtex4}
\def\mysection#1{{\bf #1.} }

\def\lsim{\mathrel{\rlap{\lower4pt\hbox{\hskip1pt$\sim$}}
    \raise1pt\hbox{$<$}}}         
\def\gsim{\mathrel{\rlap{\lower4pt\hbox{\hskip1pt$\sim$}}
    \raise1pt\hbox{$>$}}}         
\begin{document}
\begin{titlepage}

\vskip -.6in
\flushleft  \vspace*{-1.5cm}{\small BNL-HET-04/5, LBNL-55324}
\vskip .35in
\begin{center}
{\Large \bf B-factory Signals for a Warped Extra Dimension}
\vskip .17in
{\bf Kaustubh Agashe}$^*$\footnote{email: kagashe@pha.jhu.edu}, 
{\bf Gilad Perez}$^\dagger$\footnote{email: gperez@lbl.gov}
and {\bf Amarjit Soni}$^\#$\footnote{email: soni@quark.phy.bnl.gov}  
\vskip .22in
{\em $^*$ Department of Physics and Astronomy,
Johns Hopkins University, \\
Baltimore, MD 21218-2686} 
\vskip .08in
 {\em $^\dagger$ Theoretical Physics Group, Lawrence Berkeley National Laboratory,\\ Berkeley, CA 94720}                 
\vskip .08in
{\em $^\#$ Brookhaven National Laboratory\, \\
Upton, NY 11973}

\end{center}
\vskip .04in

\begin{center} {\bf Abstract}\\\end{center}
We study predictions for B-physics in a class of models,
recently introduced,
with a non-supersymmetric warped extra dimension.
In these models few ($\sim 3$)
 TeV Kaluza-Klein
 masses are consistent with electroweak data due to bulk custodial
 symmetry.
Furthermore, there is an analog of GIM mechanism
which is violated by the heavy top quark (just as in SM)
leading to striking 
signals at $B$-factories:
 (i) New Physics (NP)
contributions to $\Delta F= 2$ transitions are comparable to SM. This implies
that, within this NP framework, the success of the SM unitarity triangle fit is 
a ``coincidence''. Thus, clean extractions of unitarity angles
via e.g. $B \rightarrow \pi \pi,\rho \pi, \rho \rho, DK$
are likely to be affected, in addition to $O(1)$ deviation
from SM prediction in $B_s$ mixing.
(ii) $O(1)$ deviation from SM predictions for 
 $B \to X_s l^+ l^-$
in rate as well as in  
forward-backward and direct CP asymmetry.
(iii) Large 
 mixing-induced CP asymmetry in radiative $B$ decays, wherein the
 SM unamibgously predicts very small asymmetries. 
Also, with KK masses 3 TeV or less, 
and with anarchic Yukawa masses,  
contributions to electric dipole moments of the neutron are 
roughly 20 times larger
than the current experimental bound 
so that this framework has a ``CP problem''.
\vskip .2in

\end{titlepage}
\newpage
\renewcommand{\thepage}{\arabic{page}}
\setcounter{page}{1}

\mysection{Introduction}
The standard model (SM) 
is in very
good agreement with data. 
However, it is widely perceived to be an incomplete theory.
In particular, in the SM, the hierarchy between the Planck scale and
the electroweak symmetry breaking (EWSB) scale, $\Lambda_{\rm EWB}$,
is unnatural since the Higgs mass is ultra-violet (UV) sensitive.

Solutions to the hierarchy problem, therefore, involve extending the
SM at just above 
the EWSB scale which, in
general, 
spoils the good agreement of the SM with data, for example, by introducing
new sources of flavor violation. Given this inherent tension, it is important to
identify new physics [NP] frameworks which preserve the 
appealing features of the SM. 

We consider the Randall-Sundrum scenario (RS1) \cite{rs1}, which provides 
an elegant
solution to the hierarchy problem, and 
show that this framework
with mass scale of NP as low as a few TeV 
leads to many striking signals in B-physics.  
This renders B-facilities a valuable probe of the parameter 
space of this novel physics scenario.

In this framework, 
due to warped higher-dimensional spacetime, the
mass scales in an effective $4D$ description
depend on location in extra dimension:
the Higgs sector is localized at the ``TeV'' brane where it is protected
by a low warped-down fundamental scale of order a TeV while $4D$ gravity
is localized near the ``Planck'' brane which has a Planckian fundamental scale.

In the original RS1 model, the entire SM was localized 
on the TeV brane. In this set-up, 
flavor issues are sensitive to the UV completion of the RS1
effective field theory: there is no
understanding of hierarchies in fermion masses or of
smallness of flavor changing neutral currents
(FCNC) from higher-dimensional operators which are naively
too large being suppressed only by the warped-down cut-off $\sim$ TeV.

Allowing the SM fermions and gauge fields to propagate in the
bulk makes flavor issues UV-insensitive as follows. The 
light fermions can be localized near the Planck brane
(using a $5D$ fermion mass parameter \cite{gn, gp}) where the effective cut-off 
is much higher than
TeV
so that FCNC's from higher-dimensional operators
are suppressed \cite{gp, hs}. Moreover, this results in small $4D$
Yukawa couplings to the Higgs,
even if there are no 
small $5D$ Yukawa couplings \cite{gp, hs}. 
The top quark can be localized near TeV brane
to obtain a large $4D$ top Yukawa coupling.
This provides an understanding of
hierarchy of fermion masses (and mixing) with{\em out} hierarchies in 
fundamental ($5D$) parameters solving
the SM flavor puzzle.

With bulk fermions and gauge fields, 
FCNC's from exchange of gauge Kaluza-Klein (KK) modes are induced. 
The couplings of light
fermions to gauge KK modes are 
flavor dependent
(due to different wave-functions of fermions in the 5D), but, remarkably, 
this flavor dependence is small~\cite{universal},  
since gauge KK modes are localized near the TeV brane,
whereas light fermions are localized near the Planck brane
(unlike the {\em flat} extra dimension case). Thus, FCNC's 
for light fermions from exchange of gauge KK modes are suppressed \cite{gp, hs}.    
Recall that this is the same reason why these fermions are light,
i.e., the Higgs is also
localized near TeV brane just like the gauge KK modes. Thus, this RS1 model has
{\em built-in}  
analog of the SM GIM mechanism and approximate
flavor symmetries for the light fermions~\cite{future}. 

In spite of all these appealing features the above
framework has another source of tension. 
It was shown that in order to 
be consistent with 
electroweak precision measurements
(EWPM)
one needs to assume $m_{\rm KK}\gsim 10\,$TeV~\cite{EWPM}.
This brings a little hierarchy problem related to the
smallness of $\left({\Lambda_{\rm EWB}/ m_{\rm KK}}\right)^2 \sim10^{-4},\,$ 
which is radiatively unstable.

Recently, it was shown
that,
with enhanced bulk
electroweak (EW) gauge symmetry,
$SU(2)_L \times SU(2)_R \times U(1)_{ B - L }$, 
few TeV KK masses are consistent with EWPM~\cite{custodial}. 
This gauge structure, 
with custodial symmetry, 
in warped
space has also been used to construct Higgsless models of EWSB in~\cite{csaki,Nomura}. 

Our goal is to systematically study the flavor structure of the above framework
and to re-investigate the NP contribution to FCNC processes~\cite{hs}
in the presence of rather light KK excitations.
In this Letter, we will summarize the main
effects and the ramifications for B-factories~\cite{future}.
Even though we mainly concentrate on RS1 model with the Higgs strictly
localized on TeV brane, 
our results will be valid for warped space models with{\em out} the Higgs or
with 5D profile for Higgs as well.
Thus these results are model independent within the new class of models which
assume RS1 with custodial gauge symmetry in the bulk!

\mysection{Flavor violation}
Most of the flavor-violating effects
are related to the violation of RS-GIM mechanism by large top quark mass 
(just as in SM) as follows.
$(t,b)_L$ cannot be localized near the Planck brane. 
This is since the large top mass then requires a 
very large
$5D$ Yukawa coupling such that the theory is 
strongly coupled at the scale of the first KK mode~\cite{iso}.
Thus,
$(t,b)_L$ must be localized 
near the TeV brane.
This has two consequences:
(1) In the interaction basis, the coupling of $b_L$ to gauge KK modes (say the gluons),
$g_{ G^{ \rm KK } }^b$, is 
large compared with the ones of the lighter quarks.
This is a source of flavor violation which
combined with mixing yields FCNC processes.
(2) The Higgs VEV mixes zero and KK modes of $Z$ leading to a
non-universal shift, in the interaction basis, of the coupling of
$b_L$ to the physical $Z$~\cite{custodial, bn}:  
\begin{eqnarray}
\delta g_Z^{ b} & \sim & g_{Z^{\rm KK }}^{b} 
\sqrt{ \log \left( M_{ Pl } / \hbox{ TeV } \right) }
\frac{ m_Z^2 }{ m_{\rm KK }^2 }
\label{Zbb}\,,
\end{eqnarray}
where $g_{Z^{\rm KK }}^{b}$ is the (non-universal) coupling 
between $b_L$ and a KK $Z$ state
before EWSB.
The factor of
$\sqrt{ \log \left( M_{ Pl } / \hbox{ TeV } \right) }$ comes from enhanced
Higgs coupling to gauge KK mode (since they are 
localized near the TeV brane). 
EWPM, which are not related to flavor mixing, require that 
this shift is smaller than $\sim 1 \%$.
Note that in the above framework the interaction basis is unique. This
is the basis in which all the interactions (except Yukawa
ones), including ones with
KK modes are flavor diagonal.

There is, therefore, a tension
between obtaining a large top Yukawa and not introducing a too large
flavor violation in processes related to EWPM~\cite{custodial, bn}.
As a result, all the models assume the following:
(1) A large (close  to maximal) dimensionless $5D$ Yukawa, $\lambda_{
  5D } \sim 4$
(such that the weakly coupled effective theory contains
3-4 KK modes). 
(2) The wave function of $(t_L,b_L)$ is localized as close  
to the TeV
brane 
as is allowed by
EWPM
so that a large (close to maximal) shift in coupling of $b_L$ to $Z$,  
$\delta g_Z^{ b}\sim 1 \%$ is obtained.
Using Eq. (\ref{Zbb}) we can summarize the above by the following~\cite{future}
\begin{eqnarray}
  g_{ G^{ \rm KK } }^b & \sim & g_s\,, \ \ \
  g^b_{Z^{\rm KK} } \sim g_Z\,, \ \ \   m_{ KK } \sim 3 \, \hbox{TeV}
\label{gbb}.
\end{eqnarray}
This (unavoidable) setup leads to 
sizable NP contributions in the following three kinds of top quark dominated FCNC processes:
(i) $\Delta F = 2$; (ii) $\Delta F = 1$ governed by box and EW penguin
diagrams; (iii)
radiative decays. We will consider these effects, in turn, in what follows.

Let us elaborate more on how the NP contributions arise in the above framework.
Since the couplings of fermions of a given type to gauge KK modes
are non-universal, flavor mixing is induced when  
a transformation from the (unique) interaction to the
mass basis is performed for the quarks.
Here, we mostly consider couplings of {\em left}-handed down
quarks. This is since
coupling of $b_L$ to gauge KK modes is abnormally large~\cite{burdman}
compared with the couplings of other down type quarks (which are
localized near the Planck brane):
the coupling of $b_L$ to gauge KK modes
is ``dictated'' by $m_t$.
Consequently a large RS-GIM violation is induced.

\mysection{Signals}
To study flavor-changing effects, we 
need to estimate the mixing angles of the unitary transformations.
Generically  5D Yukawa matrices are expected to be anarchic. 
Thus the mixing angles of the transformations 
are roughly given by ratio of wave-functions at TeV brane.
The unitary transformations 
for left-handed up, $U_L$, and down quarks, $D_L$,
are similar due to the fact that $u_L$ and $d_L$ have the same wave-function. 
Since $U_L D^{ \dagger }_L = V_{\rm CKM }$, we get
$D_L \sim V_{\rm CKM }$. It follows, therefore, that the gluon 
KK mode-$b_L$-$(s, d)_L$ vertex is roughly given by $g_{ G^{ \rm KK }}^b 
V_{ t (s, d) }$ while 
the gluon KK mode-$d_L$-$s_L$ one is $g_{ G^{ \rm KK } }^b V_{ ts } V_{ td }$.

This results in {\em tree}-level exchange of gluon KK mode contributing
to $\Delta F = 2$ operators: 
\begin{eqnarray}
  \frac{ M_{ 12 }^{\rm RS} }{ M_{ 12 }^{\rm SM} }
 \hspace{-.035cm}  \sim \hspace{-.035cm} 16 \pi^2\,\frac{ \left(g^b_{ G^{ \rm KK } } \right)^2  }{ g_2^4}\,
  \frac{ m_W^2  }{ m_{\rm KK }^2 } 
\sim C \left(g^b_{ G^{ \rm KK } } \right)^2
\left( \frac{ 3 \hbox{TeV} }{ m_{\rm KK } } \right)^2 \hspace{-.04cm},
\label{F2}
\end{eqnarray}
where $C$ is an order one complex coefficient, 
mixing angles are of the same size in both RS1 and SM contributions and
$M_{12}^{\rm SM,\,RS}$ is the SM (box diagram) and RS1 (KK gluon exchange) $\Delta F=2$
transition amplitudes respectively.
We see that the KK gluon contribution to $B_d^0 - \bar{B}^0_d$,
$B_s^0 - \bar{B}^0_s$ mass difference,
$\epsilon_K$ and the CP asymmetry in $B\to \psi K_S$ is
comparable to SM ones.

The 
SM predictions depend on
$V_{ td }$ which is currently not severely constrained 
by tree-level 
decays and unitarity~\cite{pdg} which are not affected by NP
contributions.
The
data, therefore, can be fitted even with RS1 contributions 
comparable to SM~\cite{future}.

This, however, leads to a ``coincidence problem'': why is SM fit (usually
presented as a plot of the constraints in the $\rho-\eta$ plane, see 
{\it e.g.}~\cite{pdg}) so good? At present, this problem
is not so severe since there are ${\cal O} \left( 20 \% \right)$ uncertainties in
SM predictions for $\epsilon_K$ and $\Delta m _{ B_d }$ (due mainly to
hadronic matrix elements)~\cite{ENP} and also the RS1 
contributions have $O(1)$ uncertainties due
to fluctuations in $\lambda_{ 5D }$. Consequently,
clean measurements of $\alpha$ and $\gamma$ via
$B \rightarrow \pi \pi, \pi \rho, \rho \rho, DK$  
are likely to be affected.

The case of $B_s - \bar{B}_s$ mixing is slightly different than 
$B_d - \bar{B}_d$ mixing and $\epsilon_K$ 
as
the SM contribution 
is known (up to
hadronic matrix elements) 
since $V_{ ts }$
is constrained by unitarity and tree-level decays \cite{pdg}. 
Hence, for a generic order one complex coefficient (it is complex due to physical
phases in $D_L$~\cite{future}), we expect an ${\cal O}(1)$ deviation
from SM prediction in $\Delta m_{ B_s}$
(see reference \cite{burdman} for a larger effect). Similarly, 
an ${\cal O}(1)$ time-dependent CP asymmetry in 
$B_s \rightarrow J / \psi \phi$ is induced compared with the SM
${\cal O} \left( \lambda_c^2 \right)$ prediction, where
$\lambda_c\sim 0.22$.
Also deviations from SM expectation for $\gamma$ ought to occur
in $B_s \to D K$.

Next, we consider $\Delta F = 1$ transitions.
We start with the discussion of processes which in the SM
are dominated by QCD penguin diagrams such as
$b \rightarrow s \bar{s} s$. There 
is a contribution from KK gluon exchange as in the $\Delta F=2$ case.
The coupling of KK gluon to
$s$ is suppressed by
$\sim \sqrt{ \log \left( M_{ Pl } / \hbox{TeV} \right) }$
since the strange quark is localized near Planck brane
whereas the KK gluon is localized near TeV brane (this is the universal part of
coupling of fermions to gauge KK modes). Thus, it is clear that
KK gluon contribution $\sim 1/5$ SM QCD penguin. 
In addition, there is
dilution of NP effect in QCD penguin after RG scaling from
TeV to $m_b$. So, KK gluon NP contributions in
$\Delta F = 1$ transition cannot compete with SM QCD penguin ones
(see however~\cite{burdman} for a 
$O(1)$ 
effect with 
$\sim$ 1 TeV KK masses).

The contribution from exchange of {\em KK} mode of $Z$ is smaller than that from KK gluon.
However, as mentioned above, 
the KK mode of $Z$ mixes with {\em zero}-mode of $Z$ due to Higgs vev,
in turn, generating a flavor-violating coupling to the physical $Z$. Thus, we get 
the following
contributions to coefficients of 
EW
penguin operators, $C_{ 7-10 }$ (four quark)
and $C_{ 9V, \; 10 A }$ (leptonic operators): 
\begin{eqnarray}
\frac{ C^{ Z,{\rm RS} }_{ 7-10, \; 9V, \; 10A } }{ C^{ Z, {\rm SM}}_{ 7 - 10, \;
9V, \; 10A } } 
& \sim & \frac{ 16 \pi^2 }{ g_2^2 }  
\frac{ g^b_{Z^{\rm KK} }}{ g_Z } \sqrt{ \log \left( M_{ Pl} / \hbox{TeV} \right) } 
\frac{ m_Z^2 }{ m_{\rm KK }^2 } 
\nonumber \\
 & \sim & \frac{ g^b_{Z^{\rm KK} }}{ g_Z } \left( \frac{ 3 \hbox{TeV} }{ m_{\rm KK } } \right)^2\,,
\end{eqnarray}
where
superscript 
$Z$ on $C_{ 7-10, \; 9V, \; 10A }$ denotes 
$Z$ penguin part 
and, as for $\Delta F = 2$ case, the SM contribution is from
top 
quark loop
and mixing angles are of same size in both contributions.
Thus, the two contributions are comparable.
This leads to an 
${\cal O}(1)$ effect in BR for rare flavor-changing decays, for example, 
$b \rightarrow s l^+ l^-$~\cite{workshop}\cite{bn}, where uncertainty
in theory prediction is of ${\cal O} \left( 20 \% \right)$ and 
current experimental error (from BABAR and BELLE) is of ${\cal O}
\left( 30 \% \right)$~\cite{hurth}.
In addition a smoking gun signal is that 
significant departure from SM
is expected in the direct CP asymmetry and in the
spectrum of leptons in this decay,
as well as in the forward-backward 
asymmetry since
the new physics effect is only in the $Z$ penguin 
(with almost axial couplings of leptons, i.e., in $C_{ 10 A }$) and not in the
photon penguin (with vector couplings of leptons, i.e., in $C_{ 9 V }$). 

As mentioned above, the $b \rightarrow s \bar{s} s$
transition is dominated by SM QCD penguin. 
Similar RS1 contribution from flavor-violating $Z$ vertex 
is suppressed by at least $\sim g_Z^2 / g_s^2 \sim 20 \%$ and
therefore subleading~\cite{workshop}\cite{bn}.
Consequently, RS1 can accomodate only mild deviations\cite{future} 
(unlike\cite{burdman}, as mentioned above) from 
the SM in
time dependent CP asymmetry in ``penguin-doiminated" B decays,
such as $\phi (\eta', \pi^0, \omega, \rho^0) K_s$.

We next consider radiative decay processes.
Since these require helicity flip, related NP contributions appear only at the loop level
in our framework. 
The dominant contribution comes from
loops of Higgs and {\em KK} fermions since  
couplings of 
KK fermions to Higgs are enhanced. We show elsewhere~\cite{future} that
KK gluon contribution is aligned in flavor space with fermion mass matrix
and hence is not flavor-violating. We find
the following contribution to dipole operator for $b \rightarrow s$ transition:
\begin{eqnarray}
\frac{ C_{ 7 \; \gamma }^{ \prime \; {\rm RS} } }{ C_{ 7 \gamma }^{
    {\rm SM } } } & \sim
& \frac{ \lambda_{ 5D }^2 }{ g_2^2 } \frac{ m_W^2}{ m_{ KK }^2 } 
\frac{ \left( D_R \right)_{ 23 } } 
{ V_{ ts } } 
\nonumber \\
 & \sim & \left( D_R \right)_{ 23 } \left( \frac{ \lambda_{ 5D } }{4} \right)^2
\left( \frac{ 3 \hbox{TeV} }{ m_{ KK } } \right)^2\,,
\label{dipole}
\end{eqnarray}
where $C_{ 7 \; \gamma }$ and $C^{ \prime }_{ 7 \; \gamma}$ 
are coefficients of dipole operators with
$b_R$ and $b_L$, respectively
and $\left( D_R \right)_{ 23 } \rightarrow \left( D_L \right)_{ 23 }$ for 
$C^{ RS }_{ 7 \gamma }$.
For $b \rightarrow d$ transition, $\left( D_{ L, \; R } \right)_{ 23 } \rightarrow
\left( D_{ L, \; R } \right)_{ 13 }$ and $V_{ ts } \rightarrow V_{ td }$.

Let us now estimate the right-handed (RH) down quark mixing appearing 
in the above RS1 contribution.
Due to anarchic $\lambda_{ 5D }$, the
ratio of masses are also given by
ratio of wave-functions on TeV brane (just like the mixing angles) so that:
\begin{eqnarray}
\frac{ m_s }{ m_b}\, \left( D_L \right)_{ 23 }^{-1} &
\sim&  \left( D_R \right)_{ 23 }
\sim
\frac{ m_s }{ m_b}\,V_{ts}^{-1}={\cal O}(1)\,,
\end{eqnarray}
where  we used the bottom and strange quark masses
at the $\sim$ TeV scale and also $\left( D_L \right)_{ 23 } \sim V_{ ts }$.
Similarly, we find $\left( D_R \right)_{ 13 } 
\sim 
\lambda_c$, {\it i.e.}, 
RH 
down 
quark 
mixing are much larger than left-handed.
Then, from Eq. (\ref{dipole}),
we see that RS1 contribution to $b_L \rightarrow (s, d)_R \gamma$ is comparable to
SM contribution to $b_R \rightarrow (s, d)_L 
\gamma$ for $\lambda_{ 5D } \sim 4\,.$ 
Also, NP contribution to $b_R \rightarrow (s, d)_L \gamma$ is negligible
(see reference \cite{Kim:2002kk} for an earlier study of only this operator).

This leads to another smoking gun signal: ${\cal O}(1)$
mixing induced CPV due to interference between
the SM amplitude for $b_R \rightarrow (s,d)_L \gamma$ and
the NP
contribution to $b_L \rightarrow (s,d)_R \gamma$~\cite{soni} and also
deviation from a pure left handed polarization of the emitted photon~\cite{Gross}.
This will be tested in 
$B \rightarrow K^{ \ast } \gamma$, $B_s \rightarrow \phi \gamma$
($b \rightarrow s$) transitions and
$B \rightarrow \rho \gamma$,
$B_s \rightarrow K^{ \ast } \gamma$ 
($b \rightarrow d$) transitions. 

Finally, we discuss 
contribution to neutron's electric dipole moment (EDM)  
which arise from similar diagrams.
With $O(1)$ complex phases, the contribution  
exceeds the experimental limits
by ${\cal O}(20)$. We find that while contributions
from CKM-like phases vanish at the one loop level
sizable contributions are induced by Majorana-like phases. 
Though this requires flavor mixing, even with two flavors
we find unsuppressed one loop amplitudes~\cite{future}.
Thus, our framework 
has a CP problem!

\mysection{Conclusions}
Within the RS1 framework,
localization of light fermions far from the TeV brane
leads to three virtues:
(i) Suppression of higher dimensional flavor violating operators.
(ii) Suppression of flavor violating coupling to KK excitations.
(iii) A solution to the SM flavor puzzle.
There is a built-in analog of 
GIM mechanism of the SM and approximate flavor symmetries. 
As in SM, inclusion of 
heavy top quark leads to RS-GIM violation, 
in particular, to large couplings of {\em left}-handed 
bottom to gauge KK modes, in turn, resulting in 
$O(1)$ effects in $\Delta F=2$ processes and in rare flavor- changing
decays, 
for example, $b \rightarrow s l^+ l^-$.
Also, the large $5D$ Yukawa required to obtain top mass coupled with large
RH down quark mixing leads to $O(1)$ effect in radiative $B$ decays.
These B-physics signals should be of great
relevance to B-facilities in hadronic
and in $e^+ e^-$ environments.
Finally, the above framework suffers from a CP problem.

Using the
AdS/CFT correspondence, the {\em weakly}-coupled RS1 model can be viewed as
a tool to study a purely $4D$ {\em strongly} coupled conformal Higgs sector
\cite{rscft}.
Thus, a key point of our study is that a $4D$ strongly interacting Higgs sector
can solve the flavor puzzle and have suppressed
FCNC's 
with striking signals (see Table 1) at $B$ facilities.

\mysection{Acknowledgements}
K.A.~is supported by the Leon Madansky fellowship
\& NSF Grant P420D3620414350; G.P. \& A.S.
are supported by 
DOE under Contract Nos. DE-AC0376SF00098 \& 
DE-AC02-98CH10886. 
We thank
G.~Burdman, Z.~Chacko, W.~Goldberger, 
Y.~Grossman, I.~Hinchliffe, D.~E.~Kaplan,
Y.~Nir, Y.~Nomura, F.~Petriello, R.~Sundrum \& M.~Suzuki 
for discussions.

\newpage

\begin{table}
\hspace{0.5 in}
\vspace{0.2 in}
\begin{tabular}{|c|c|c|c|c|c|c|}
\hline
& $\Delta m_{B_s}$ & $S_{B_s \rightarrow \psi \phi}$ & $S_{B_d
  \rightarrow  \phi K_s}$ & $Br[b \rightarrow  s l^+ l^-]$ & $S_{B_{d,s} \rightarrow  K^*,\phi
  \gamma}$ &
$S_{B_{d,s} \rightarrow  \rho,K^*\gamma}$
\\
\hline
\hline
RS1 & $\Delta m_{B_s}^{\rm SM}[1+O(1)]$ & $O(1)$ & $\sin 2\beta \pm
O(.2)$ & $Br^{\rm SM}[1+O(1)]$ & $O(1)$ & $O(1)$ \\
\hline
SM & $\Delta m_{B_s}^{\rm SM}$ & $\lambda_c^2$ & $\sin 2\beta$ &
$Br^{\rm SM}$ & ${m_s\over m_b} \left(\sin 2
\beta,\lambda_c^2\right)$ & ${m_d\over m_b}\left(\lambda_c^2,\sin2\beta\right)$ \\
\hline
\end{tabular}
\caption{Contrasting signals from RS1 with the SM}
\label{RS1_tab}
\end{table}
                                                                                


\begin{thebibliography}{99}

\bibitem{rs1}L.~Randall and R.~Sundrum, 
Phys.\ Rev.\ Lett.\  {\bf 83}, 3370 (1999)
[arXiv:hep-ph/9905221].


\bibitem{gn}Y.~Grossman and M.~Neubert, 
Phys.\ Lett.\ B {\bf 474}, 361 (2000)
[arXiv:hep-ph/9912408].

\bibitem{gp}T.~Gherghetta
and A.~Pomarol,
Nucl.\ Phys.\ B {\bf 586}, 141 (2000)
[arXiv:hep-ph/0003129].

\bibitem{hs}For earlier work 
assuming $\sim 10$ TeV KK masses, see: S.~J.~Huber and Q.~Shafi, 
Phys.\ Lett.\ B {\bf 498}, 256 (2001)
[arXiv:hep-ph/0010195];
G.~Burdman,
Phys.\ Rev.\ D {\bf 66}, 076003 (2002)
[arXiv:hep-ph/0205329];
S.~J.~Huber,
Nucl.\ Phys.\ B {\bf 666}, 269 (2003)
[arXiv:hep-ph/0303183];
S.~Khalil and R.~Mohapatra,
arXiv:hep-ph/0402225.


\bibitem{universal} There is a sizable flavor-universal part of
this coupling.


\bibitem{future}For further details
see, K.~Agashe, G.~Perez and A.~Soni, in preparation. 



\bibitem{EWPM}S.~J.~Huber and Q.~Shafi,
Phys.\ Rev.\ D {\bf 63}, 045010 (2001)
[arXiv:hep-ph/0005286];
S.~J.~Huber, C.~A.~Lee and Q.~Shafi,
Phys.\ Lett.\ B {\bf 531}, 112 (2002)
[arXiv:hep-ph/0111465];
C.~Csaki, J.~Erlich and J.~Terning,
Phys.\ Rev.\ D {\bf 66}, 064021 (2002)
[arXiv:hep-ph/0203034];
J.~L.~Hewett, F.~J.~Petriello and T.~G.~Rizzo,
JHEP {\bf 0209}, 030 (2002)
[arXiv:hep-ph/0203091].



\bibitem{custodial}K.~Agashe {\it et al.},  
JHEP {\bf 0308}, 050 (2003)
[arXiv:hep-ph/0308036].



\bibitem{csaki}C.~Csaki {\it et al.}, 
Phys.\ Rev.\ Lett.\  {\bf 92}, 101802 (2004)
[arXiv:hep-ph/0308038] and
arXiv:hep-ph/0310355.


\bibitem{Nomura}Y.~Nomura,
JHEP {\bf 0311}, 050 (2003)
[arXiv:hep-ph/0309189].



\bibitem{iso} There is also a limit on how close $t_R$ can be to 
the TeV brane since its
custodial isospin partner becomes too 
light which leads to phenomenological problems.
Thus this cannot be used to enhance the top 
Yukawa coupling~\cite{custodial}.


\bibitem{bn}G.~Burdman and Y.~Nomura,
arXiv:hep-ph/0312247.


\bibitem{burdman}For an earlier RS1 study with low KK masses, see
G.~Burdman,
Phys. \ Lett.\ B {\bf 590}, 86 (2004)[arXiv:hep-ph/0310144].
It differs from our work since hierarchies in the $5D$ Yukawas were
allowed and
constraints from $Z\to b\bar b$ were {\em not} correlated 
with the FCNC analysis.

\bibitem{pdg}K.~Hagiwara {\it et al.}  [Particle Data Group Collaboration],
Phys.\ Rev.\ D {\bf 66}, 010001 (2002).

\bibitem{ENP}
G.~Eyal, Y.~Nir and G.~Perez,
JHEP {\bf 0008}, 028 (2000)
[arXiv:hep-ph/0008009].



\bibitem{workshop}K.~Agashe, Workshop on  
a B Factory at $10^{ 36 }$ Luminosity,
SLAC, (Oct, 2003) [to appear in proceedings].


\bibitem{hurth}T.~Hurth,
Rev.\ Mod.\ Phys.\  {\bf 75}, 1159 (2003)
[arXiv:hep-ph/0212304];
A.~Ghinculov {\it et al.}, 
Nucl.\ Phys.\ B {\bf 685}, 351 (2004)
[arXiv:hep-ph/0312128];
for a recent experimental review see: M.~Nakao, arXiv:hep-ex/0312041. 







\bibitem{Kim:2002kk}
C.~S.~Kim {\it et al.},
Phys.\ Rev.\ D {\bf 67}, 015001 (2003)
[arXiv:hep-ph/0204002].

\bibitem{soni}D.~Atwood, M.~Gronau and A.~Soni,
Phys.\ Rev.\ Lett.\  {\bf 79}, 185 (1997)
[arXiv:hep-ph/9704272];
B.~Aubert {\it et al.} [BABAR Collaboration], arXiv:hep-ex/0405082. 

\bibitem{Gross}
M.~Gronau {\it et al},
Phys.\ Rev.\ Lett.\  {\bf 88}, 051802 (2002)
[arXiv:hep-ph/0107254].



\bibitem{rscft}N.~Arkani-Hamed, M.~Porrati and L.~Randall,  
JHEP {\bf 0108}, 017 (2001)
[arXiv:hep-th/0012148].


\end{thebibliography}
\end{document}